\documentclass[12pt]{article}

\usepackage{amsfonts,amsmath,amssymb}
\usepackage{hyperref}
\usepackage{cite}
\usepackage{epsfig}
\usepackage{paralist}
\usepackage{fancyhdr}
\usepackage{tikz}
\usepackage{tkz-euclide}
\usetikzlibrary{decorations.pathmorphing,patterns,calc,snakes,arrows}

\usepackage{graphicx}
\usepackage{xcolor}
\numberwithin{equation}{section}
\usepackage{etex}
\usepackage{braket}
\usepackage{float}

\def\spa#1{\phantom{\fbox{\rule[-#1cm]{0cm}{0cm}}}}

\def\be{\begin{equation}}
\def\ee{\end{equation}}
\def\bea{\begin{eqnarray}}
\def\eea{\end{eqnarray}}

\def\Tr{\mbox{Tr}}
\def\del{\partial}

\renewcommand{\thefootnote}{\fnsymbol{footnote}}

\makeatletter
\g@addto@macro\bfseries{\boldmath}
\makeatother

\def\zb{{\bar{z}}}

\def\Tr{\mathop{\mathrm{Tr}}\nolimits}

\begin{document}

\hfuzz=100pt
\title{{\Large \bf{Replica Wormholes from Liouville Theory}}}
\date{}
\author{Shinji Hirano$^{a, c}$\footnote{
	e-mail:
	\href{mailto:shinji.hirano@wits.ac.za}{shinji.hirano@wits.ac.za}}
	~and Tsunehide Kuroki$^{b}$\footnote{
	e-mail:
	\href{mailto:kuroki@toyota-ti.ac.jp}{kuroki@toyota-ti.ac.jp}} 	
}
\date{}

\maketitle

\thispagestyle{fancy}
\rhead{YITP-21-100}
\cfoot{}
\renewcommand{\headrulewidth}{0.0pt}

\vspace*{-1cm}
\begin{center}
$^{a}${{\it School of Physics and Mandelstam Institute for Theoretical Physics }}
\\ {{\it University of the Witwatersrand}}
\\ {{\it 1 Jan Smuts Ave, Johannesburg 2000, South Africa}}
  \spa{0.5} \\
$^b${{\it Theoretical Physics Laboratory, Toyota Technological Institute}}
\\ {{\it 2-12-1 Hisakata, Tempaku-ku, Nagoya 468-8511, Japan}}
\spa{0.5}  \\
\&
\spa{0.5}  \\
$^c${{\it Center for Gravitational Physics}}
\\ {{\it  Yukawa Institute for Theoretical Physics, Kyoto University}}
\\ {{\it Kitashirakawa-Oiwakecho, Sakyo-ku, Kyoto 606-8502, Japan}}
\spa{0.5}  

\end{center}

\begin{abstract}

The replica wormholes are a key to the existence of the islands that play a central role in a recent proposal for the resolution of the black hole information paradox.
In this paper, we study the replica wormholes in the JT gravity, a model of two-dimensional quantum gravity coupled to a non-dynamical dilaton, by making use of the 2$d$ conformal field theory (CFT) description, namely, the Liouville theory coupled to the $(2,p)$ minimal matter in the $p\to\infty$ limit. 
In the Liouville CFT description, the replica wormholes are created by the twist operators and the gravitational part of the bulk entanglement entropy can be reproduced from the twist operator correlators. We propose the precise dictionary and show how this correspondence works in detail.   

\end{abstract}

\renewcommand{\thefootnote}{\arabic{footnote}}
\setcounter{footnote}{0}

\newpage

\tableofcontents


\section{Introduction}
\label{Sec:Introduction}

Over the last couple of years, there has been significant progress towards the resolution of the black hole information paradox \cite{Penington:2019npb, Almheiri:2020cfm, Almheiri:2019psf, Almheiri:2019hni, Almheiri:2019yqk}. 
Instrumental to this new development is a simple model of quantum black holes and the Hawking radiation based on the JT gravity, a model of two-dimensional quantum gravity coupled to a non-dynamical dilaton  \cite{Jackiw:1984je, Teitelboim:1983ux}. 
Very importantly, the JT gravity is not merely a toy model of quantum black holes but it also provides a universal low temperature description for near-extremal charged and rotating black holes in four and five dimensions \cite{Nayak:2018qej, Moitra:2018jqs, Moitra:2019bub}. We can thus hope to learn a general lesson for the information paradox that can, in principle, be applied to semi-realistic black holes in our universe by further studying the JT gravity. 
 
\begin{figure}[h!]
\centering
\centering \includegraphics[height=1.5in]{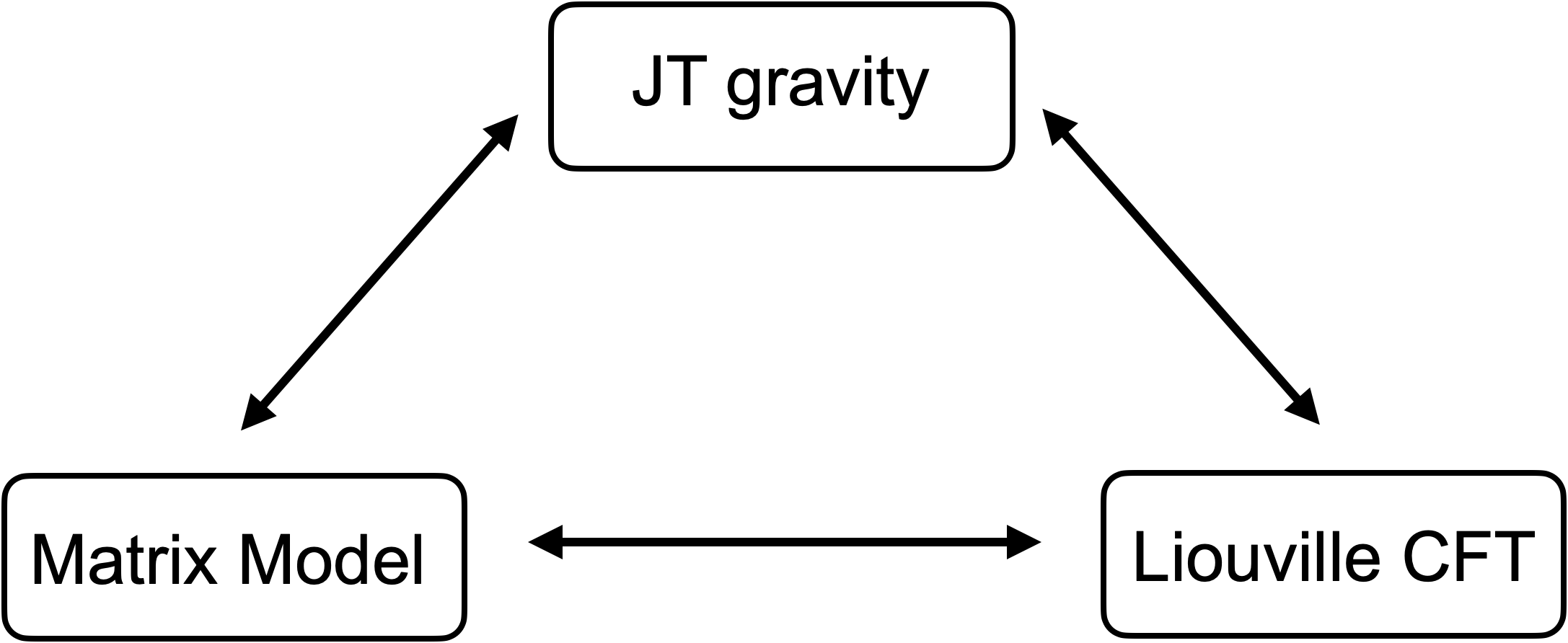}
\vspace{-0cm}
\caption{The JT gravity triality: There are at least three descriptions of the JT gravity. The first is the two-dimensional quantum gravity coupled to a non-dynamical dilaton. This is the so-called JT gravity first studied by Jackiew and Teitelboim \cite{Jackiw:1984je, Teitelboim:1983ux}. The second is a certain double-scaled matrix model proposed by Saad, Shenker, and Stanford (SSS) \cite{Saad:2019lba}. The third is the Liouville CFT that is a two-dimensional quantum gravity coupled to a minimal model and the continuum description of the SSS matrix model.}
\label{fig:JTtriality}
\end{figure}  
The JT gravity enjoys at least three different descriptions as illustrated in Figure \ref{fig:JTtriality}: 
It is a model of two-dimensional quantum gravity coupled to a dilaton defined by the action \eqref{eq:JTaction}
as originally proposed by Jackiw and Teitelboim \cite{Jackiw:1984je, Teitelboim:1983ux}.
The second description is via a hermitian one-matrix model in the double scaling limit as proposed by Saad, Shenker, and Stanford (SSS) \cite{Saad:2019lba}. 
The third is the Liouville field theory coupled to conformal matter in the large central charge limit.
The connection between the latter two was originally suggested by SSS in \cite{Saad:2019lba} and further developed in \cite{Mertens:2019tcm, Mertens:2020hbs, Gregori:2021tvs, Turiaci:2020fjj,Okuyama:2021eju}.
In hindsight, the equivalence of the latter two is no surprise and quite natural since it has been well established that the Liouville CFT or minimal string theory is a continuum description of the (old) matrix models \cite{Ginsparg:1993is}. Furthermore, the more precise relation of the SSS matrix model to the old matrix models was established in \cite{Okuyama:2019xbv} and exploited to find the non-perturbative completion of the SSS matrix model in \cite{Johnson:2019eik}.
In this paper, we focus more on the direct connection between the first and the third building on the relation between the second and the third.\footnote{More recently, an alternative proposal was made to connect the JT gravity and the Liouville CFT \cite{Suzuki:2021zbe}. We do not have a clear understanding of how the two versions of the JT gravity/Liouville CFT correspondence are related. It is also worth mentioning that in a somewhat related work \cite{Betzios:2020nry}, the $c=1$ and minimal string theories were studied from the perspective of the factorization problem \cite{Saad:2019lba} in the JT gravity.}

From the perspective of  black hole physics, it is of particular importance to understand how the results on the black hole information in the JT gravity  can be understood in the Liouville CFT description. In particular, the replica wormholes \cite{Penington:2019kki, Almheiri:2019qdq} turned out to be a key to the existence of the islands that play a central role in a recent proposal for the resolution of the black hole information paradox \cite{Penington:2019npb, Almheiri:2020cfm, Almheiri:2019psf, Almheiri:2019hni, Almheiri:2019yqk}.
So our goal in this paper is to provide a new description of replica wormholes in terms of the Liouville CFT. 
As we will see, they are created by the twist operators in the Liouville CFT and the gravitational part of the bulk entanglement entropy \cite{Almheiri:2019qdq} can be reproduced from the twist operator correlators with a certain prescription.

The organization of the paper is as follows: In Section \ref{Sec:JTgL}, we give a brief review on the basics of the correspondence between the JT gravity and the Liouville CFT.
Along the way, by studying the disk partition function in detail, we provide our own account of the dictionary between the two and add some new observations.
In Section \ref{Sec:RW}, we present our main results, namely, the Liouville CFT description of replica wormholes. In particular, the cosmic branes that create replica wormholes are identified with the twist operators in the Liouville CFT, and as evidence for our proposal, we reproduce the gravitational part of the bulk entanglement entropy by using the standard replica trick in 2$d$ CFT.
In Section \ref{Sec:Mdd}, we slightly extend the dictionary in Section \ref{Sec:JTgL} by including conical defects in AdS$_2$. In particular, we study a very special defect, dubbed ``marginal defect,'' in detail for which some interesting exact result can be found.
We conclude our paper with a few comments for the future in Section \ref{Sec:Discussions}.


\section{JT gravity/Liouville CFT correspondence}
\label{Sec:JTgL}

The JT gravity is a model of two-dimensional quantum gravity that couples to a non-dynamical dilaton $\phi$ described by the action \cite{Jackiw:1984je, Teitelboim:1983ux}\footnote{We adopt the convention of \cite{Saad:2019lba} and set $4G=1$. The extrinsic curvature $K=1$ receives the order ${\cal O}(z^2)$ correction that yields the Schwarzian action \eqref{Schwarzian} below. We thank Thomas Mertens for pointing out the inaccurate statement on this point in the first version of our paper.}
\begin{align}
I_{JT}=-\underbrace{{S_0\over 2\pi}\biggl[{1\over 2}\int_{\cal M}\sqrt{g}R+\int_{\partial{\cal M}}\sqrt{h}K\biggr]}_{S_0\,\times\,\chi({\cal M})}
-\underbrace{\biggl[{1\over 2}\int_{\cal M}\sqrt{g}\Phi(R+2)+\int_{\partial{\cal M}}\sqrt{h}\Phi(K-1)\biggr]}_{R\,=\,-2\quad\&\quad K\,=\,1+{\cal O}(z^2)}
\label{eq:JTaction}
\end{align}
and the further coupling to the matter.
In this paper, we focus on the pure dilaton gravity and do not consider the coupling to the matter as the dilaton gravity sector is sufficient for our purpose of studying the replica wormholes. 

The first part of the action is topological and the Euler number $\chi({\cal M})$ of a Riemann surface ${\cal M}$. So the dynamics is solely determined by the second part of the action. Since the dilaton has no kinetic term and is thus non-dynamical, the equation of motion from the $\Phi$ variation, or equivalently, the path-integral over $\Phi$, constrains the two-dimensional manifolds with boundaries to those of constant negative curvature and positive boundary extrinsic curvature, i.e., $R=-2$ and $K=1 +{\cal O}(z^2)$, respectively. In this paper, we focus on a disk topology, i.e., the Euclidean $AdS_2$ space ($EAdS_2$), or equivalently, the 2$d$ hyperbolic space $H_2$. 
Even though the dilaton $\Phi$ is non-dynamical, it plays an essential role in the boundary dynamics of the JT gravity. The equations of motion from the metric $g_{ab}$ variations yield
\begin{align}
\nabla_a\nabla_b\Phi-g_{ab}\nabla^2\Phi+g_{ab}\Phi=0\ .
\end{align}
In the Poincar\'e patch of $EAdS_2$ where $ds^2=(d\tau^2+dz^2)/z^2$, these equations, in particular, have the solution
\begin{align}\label{dilaton}
\Phi={2\pi\gamma\over z}\ .
\end{align}
This sets the boundary condition of the system. The constant $\gamma$ introduces a length scale to the theory and breaks the 1$d$ boundary conformal symmetry, which generates nontrivial dynamics described by the boundary Schwarzian action at $z\to 0$ \cite{Jensen:2016pah, Maldacena:2016upp, Engelsoy:2016xyb}:
\begin{align}\label{Schwarzian}
I_{\rm Sch}=-\gamma\int_0^{\beta} du \left\{\tau, u\right\}
\qquad\mbox{where}\qquad 
\left\{\tau, u\right\}\equiv\left({\tau''\over \tau'}\right)'-{1\over 2}\left({\tau''\over \tau'}\right)^2\ ,
\end{align}
where $\tau(u)$ is the boundary graviton corresponding to the degree of freedom of the boundary diffeomorphism $u\mapsto \tau(u)$, and we introduced the notation $\tau'\equiv d\tau/du$. The Poincar\'e $EAdS_2$ can be mapped to $H_2$ with the metric $ds^2=d\rho^2+\sinh^2\rho d\theta^2$ so that the boundary is compactified on an $S^1$. In this map, the Euclidean time $\tau$ is related to the angle $\theta$ by $\tau = \tan(\theta/2)$ as one approaches the boundary $\rho\to\infty$.\footnote{The Schwarzian derivative maps as $\{\tau,u\}=\{\tau, \theta\}\theta'^2+\{\theta,u\}={1\over 2}\theta'^2+\{\theta,u\}$. } 

In the meantime, the connection between the Liouville CFT and the JT gravity was suggested by Saad, Shenker, and Stanford (SSS) in their study of a matrix model description of the JT gravity \cite{Saad:2019lba}. The Liouville CFT can be thought of as the continuum limit of the SSS matrix model. More precisely, it is the Liouville theory coupled to the $(2,p)$ minimal matter in the $p\to\infty$ limit.  The correspondence between the two have been studied further in \cite{Mertens:2019tcm, Mertens:2020hbs}.
The Liouville CFT is defined by the action
\begin{equation}\label{LiouvilleAction}
S_L={1 \over 4\pi}\int_{\cal M} d^2x\sqrt{g}\left(g^{ab}\del_a\phi\del_b\phi
+QR\phi+\underbrace{4\pi\mu e^{2b\phi}}_{\rm bulk\,\,CC}\right)+\int_{\del{\cal M}} d\tau\sqrt{\gamma}\left(\underbrace{\mu_B e^{b\phi}}_{\rm bdy\,\,CC}+{Q\phi\over 2\pi}K\right)\ ,
\end{equation}
where the background charge $Q$ is related to the parameter $b$ by $Q=b+1/b$ and the central charge is $c_L=1+6Q^2$. The Liouville field $\phi$ appears  via conformal anomaly as the conformal mode of two-dimensional quantum gravity coupled to the conformal matter with central charge $c_m=25-6Q^2=1-3(p-2)^2/p$. So it is a model of two-dimensional quantum gravity, and one can anticipate a direct correspondence between the Liouville CFT and the JT gravity.
Since the JT gravity corresponds to the $p\to\infty$ limit, we will be most interested in the $b\to 0$ limit in which the Liouville CFT becomes semiclassical.

\begin{table}[h!]
\centering
\begin{tabular}{|c|c|}
\hline
  \multicolumn{2}{|c|}{Liouville CFT} \\
  \hline
background charge & $Q = b+1/b$ \\\hline
central charge & $c_L = 1+6Q^2$ \\\hline
bulk cosmological constant (CC) & $\mu$  \\\hline
boundary CC & $\mu_B=-\kappa-E_L$ \\\hline
boundary length & $\ell$ \\\hline
boundary ``zero point energy'' & $\kappa= \sqrt{\mu/\sin(\pi b^2)}$ \\\hline
\end{tabular}
\caption{The parameters in the Liouville CFT}\label{table:Liouville}
\end{table}

\subsection{Disk partition function}
\label{Sec:DiskPT}

Our first task is to start creating a precise dictionary between the JT gravity and the Liouville CFT \cite{Mertens:2019tcm, Mertens:2020hbs, Maxfield:2020ale}.
In particular, our goal in this section is to establish the correspondence between the parameters listed in Table \ref{table:correspondence}.
\begin{table}[h!]
\centering
\begin{tabular}{|c|c|c|}
\hline
  \multicolumn{3}{|c|}{JT gravity vs. Liouville CFT} \\
  \hline
&\quad JT gravity \quad&\quad Liouville CFT \quad\\\hline
genus counting & $S_0$ & $\ln\mu/(2b^2)$ \\\hline
boundary length to dilaton ratio& $\beta/\gamma$ & $4\pi^2\kappa\ell b^4$ \\\hline
dilaton\,$\times$\,energy & $\gamma E$ & $E_L/(4\pi^2\kappa b^4)$ \\\hline
\end{tabular}
\caption{The correspondence between the parameters of the two descriptions}\label{table:correspondence}
\end{table}
For this purpose we examine the most elementary quantity in the JT gravity, namely, the disk partition function  \cite{Saad:2019lba}
\begin{align}\label{JTdiskpartitionfunction}
Z^{\rm disk}_{\rm JT}(\beta)={e^{S_0}\over\sqrt{2\pi}}\left({\gamma\over\beta}\right)^{3/2}e^{2\pi^2\gamma\over\beta}
\end{align}
that can be computed from the boundary Schwarzian theory defined by the action \eqref{Schwarzian}.
The strategy is to reproduce it from the Liouville CFT description.

The Liouville theory on a disk is a boundary conformal field theory (BCFT). Since there is a dynamical degree of freedom $\tau(u)$ in the boundary of the JT gravity,
the Neumann boundary condition must be imposed in the Liouville CFT. It is well-known that the Neumann boundary condition in the 
Liouville theory is described by FZZT branes \cite{Fateev:2000ik, Teschner:2000md}. 
In particular, the bulk one-point function for the vertex operator $V_{\alpha}(z)\equiv e^{2\alpha\phi(z)}$ of dimension $\Delta_{\alpha}=\alpha(Q-\alpha)$ with the FZZT boundary condition is given by  \cite{Fateev:2000ik}
\begin{align}\label{FZZT1pt}
\hspace{-.3cm}
U(\alpha)\equiv\left\langle V_{\alpha}(0)\right\rangle_{\rm FZZT}
={2\over b}(\pi\mu\gamma(b^2))^{{Q-2\alpha\over 2b}}
\Gamma(2b\alpha-b^2)\Gamma\left({2\alpha\over b}-{1 \over b^2}-1\right)\cosh(2\alpha-Q)\pi s,
\end{align}
where $\gamma(b^2)=\Gamma(b^2)/\Gamma(1-b^2)$ and $s$ is a parameter related to the bulk and boundary cosmological constants
$\cosh^2(\pi bs)={\mu_B^2 \over \mu}\sin(\pi b^2)$. For later convenience, we introduce the parameter 
\be
\kappa^2\equiv{\mu\over\sin(\pi b^2)}\ .
\ee
The disk partition function corresponds to an insertion of the identity operator, {\it i.e.}, $\alpha=0$.
In the comparison between the Liouville and JT disk partition functions, however, there are a few details and subtle points to be taken into account: 
(1) The Liouville theory of our interest couples to the $(2,p)$ minimal matter in the $p\to\infty$ limit, 
so the JT limit $b=\sqrt{2/p}\to 0$ must be taken.
(2) As it will become clear in a moment, one needs to consider the ``marked'' disk partition function defined by $\partial_sU(0)$ rather than $U(0)$ itself.
Loosely speaking, the $s$-derivative corresponds to an insertion of the (integrated) boundary cosmological constant operator.
(3) The disk in the JT gravity has a fixed boundary length $\beta$. Thus, on the Liouville theory side, one needs to fix the boundary length
\be
\ell\equiv \oint_{\partial{\cal M}} d\tau e^{b\phi(\tau)}
\ee
instead of the boundary cosmological constant $\mu_B$. Hence the dictionary is given by
\begin{align}\label{diskparitionequal}
Z^{\rm disk}_{\rm L}(\ell)=Z^{\rm disk}_{\rm JT}(\beta)\ ,
\end{align} 
where $Z^{\rm disk}_{\rm L}(\ell)$ is the ``marked'' Liouville disk partition function $\times$ $Z^{\rm disk}_mZ^{\rm disk}_g$ 
and the matter and ghost disk partition functions, $Z^{\rm disk}_mZ^{\rm disk}_g$, will be normalized to $1$.
In other words, the Laplace transform of the $\partial_sU(0)$ from $\mu_B$ to $\ell$ corresponds to the JT gravity disk partition function.
(4) As it turns out, the variable conjugate to the boundary length $\ell$ is not exactly the boundary cosmological constant $\mu_B$
but rather a shifted variable 
\be\label{shiftedEL}
E_L=-\mu_B-\kappa\ .
\ee
This can be understood from the spectral representation of the disk partition function:
\begin{align}\label{diskspectral}
Z^{\rm disk}_{\rm JT}(\beta)=\int_0^{\infty}dE\rho_{\rm JT}(E)e^{-\beta E}\ ,
\end{align}
where the energy density $\rho_{\rm JT}(E)$ in the JT gravity is given by \cite{Saad:2019lba}
\begin{align}
\rho_{\rm JT}(E)=e^{S_0}{\gamma\over 2\pi^2}\sinh\left(2\pi\sqrt{2\gamma E}\right)\ .
\end{align}
So the JT disk partition function \eqref{diskspectral} must be matched to the ``marked'' Liouville disk partition function with a fixed boundary length $\ell$:
\begin{align}
Z^{\rm disk}_{\rm L}(\ell)={\cal N}\int_0^{\infty}dE_L e^{-\ell E_L}\partial_sU(0)=\int_0^{\infty}dE_L\rho_L(E_L)e^{-\ell E_L}\ ,
\end{align}
where we have introduced a normalization constant ${\cal N}$ to be fixed.
The energy density $\rho_L(E_L)$ of the Liouville theory, in the JT limit $b=\sqrt{2/p}\to 0$, reads
\begin{equation}
\begin{aligned}
\rho_L(E_L)&={\cal N}\partial_sU(0)
\simeq {\cal N}\,{2\pi\over b^2}(\pi\mu/b^2)^{{1\over 2b^2}}
\Gamma(-1/b^2)\sinh\left({1\over b^2}\cosh^{-1}(-\mu_B/\kappa)\right)\\
&\simeq {\cal N}{\cal N}'\,\mu^{{1\over 2b^2}}\sinh\left(\sqrt{2E_L/(\kappa b^4)}\right)
\end{aligned}
\end{equation}
where we have used $\pi bs=\cosh^{-1}(-\mu_B/\kappa)$  
and defined ${\cal N}'=2(\pi/b^2)^{1+1/(2b^2)}\Gamma(-1/b^2)$.
With the correspondence in Table \ref{table:correspondence},  eqn.\eqref{diskparitionequal} fixes the normalization to
\be
{\cal N}{\cal N}'={1\over 2\pi^2(4\pi^2\kappa b^4)}\ .
\ee
To be complete, we must include the $c_m=1-3(p-2)^2/p\to-\infty$ minimal matter and the $bc$ ghosts. 
However, these contributions are trivial and can be absorbed into the normalization.

\subsection{Semiclassical Liouville analysis}
\label{Sec:SCL}

In order to gain a better intuition for the JT/Liouville correspondence, we will rederive the Liouville disk partition function semiclassically from the path integral in the saddle point approximation. In fact, since the $b\to 0$ limit is the semiclassical limit of the Liouville theory, semiclassical results become exact in the JT/Liouville correspondence.
As we will see, one can attain a more direct geometric correspondence between the two theories.

To connect to the discussion in Section \ref{Sec:DiskPT}, we perform the Laplace transformation of the energy density in a slightly more general form:
\begin{align}\label{Laplace}
\int_0^{\infty} dE_Le^{-E_L\ell}\left[\nu^{-1}\sinh\left(\nu\cosh^{-1}(1+E_L/\kappa)\right)\right]
&={e^{\kappa\ell}\over\ell}K_{\nu}(\kappa\ell)\ ,
\end{align}
where we are interested in the case $\nu=1/b^2\simeq Q/b$ and we have dropped the nonessential overall constant.
The modified Bessel function on the RHS is what we expect to see in the saddle point approximation below.

We first rewrite the Liouville action \eqref{LiouvilleAction} in terms of the rescaled and shifted field $\varphi=b\phi+{1\over 2}\ln(b^2\mu)$ in the limit $b\to 0$:
\begin{equation}
\begin{aligned}\label{LiouvilleActionRescaled}
S_L=&\,\,-{\ln (b^2\mu) \over 2b^2}\chi({\cal M})+{1 \over 4\pi b^2}\int_{\cal M} d^2x\sqrt{g}\left(g^{ab}\del_a\varphi\del_b\varphi
+R\varphi+4\pi e^{2\varphi}\right)\\
&+{1\over b^2}\int_{\del{\cal M}} d\tau\sqrt{\gamma}\left({\mu_B\over\sqrt{\pi}\kappa} e^{\varphi}+{\varphi\over 2\pi}K\right)\ .
\end{aligned}
\end{equation}
This form of the action makes it clear that (1) the $b\to 0$ limit is the semiclassical limit and (2) the power of the factor $\mu^{1/(2b^2)}$ counts the Euler number of the 2-manifold ${\cal M}$. We work in the flat background $g_{z\zb}=g_{\zb z}=1/2$ with $z=re^{i\tau}$ and place the boundary at the unit circle $|z|=1$ in our analysis. 
Then, in the presence of the vertex operator $V_{\alpha}(z_i)=e^{2\alpha\phi(z_i)}$ in the bulk, the Liouville equation with the FZZT boundary condition reads
\begin{align}\label{LiouvilleEqn}
\del\bar{\del}\varphi-\pi e^{2\varphi}=-\pi b\alpha\delta^2(z-z_i)\qquad\mbox{with}\qquad
r\del_r\varphi+2\sqrt{\pi}{\mu_B\over\kappa} e^{\varphi}+1\,\biggr|_{|z|=1}=0\ .
\end{align}
In \cite{Fateev:2000ik} (see also \cite{Menotti:2006tc}) the semiclassical Liouville theory was studied in detail. In particular, the elliptic Liouville solution is given by
\be\label{ClLiouville}
e^{2b\phi}={1\over \pi\mu b^2}{a^2(1-2\eta)^2\over\left(|z|^{2\eta}-a^2|z|^{2(1-\eta)}\right)^2}\qquad\mbox{with}\qquad |z|\le 1
\quad\mbox{and}\quad 0<a<1
\ee
where we have introduced $\eta=b\alpha$ and set $z_i=0$. Note that near the location $z_i=0$ of the operator insertion, the Liouville field behaves as expected:
$b\phi\sim -2\eta\ln |z|$ with $1-2\eta>0$ and 
this yields a $\delta$-function when acted upon by the Laplacian $\del\bar{\del}$.
The FZZT boundary condition imposes a relation among the parameters:
\be\label{muBa}
-{\mu_B\over\kappa}=1+{(1-a)^2\over 2a}\qquad\Longrightarrow\qquad {E_L\over \kappa} = {(1-a)^2\over 2a}
\ee
This implies that $a\sim 1-{\cal O}(b^2)$ since $E_L/\kappa\sim{\cal O}(b^4)$ according to Table \ref{table:correspondence}. 
Moreover, as observed in \cite{Menotti:2006tc}, the boundary cosmological constant is necessarily negative, $\mu_B<0$, 
to make sense of \eqref{muBa} and thus the semiclassical limit.
It is worth noting that this corroborates the peculiar choice of sign in \eqref{shiftedEL}.

Since our aim is to study the disk partition function in the semiclassical limit, we now focus on the case of the identity operator $\eta=0$.
Recall that the Liouville CFT is a model of 2d quantum gravity in the space with the metric
\begin{align}\label{LiouvilleCLgeometry}
ds_L^2=e^{2b\phi}dzd\zb\qquad \xrightarrow{\,\, a\to 1\,\,}\qquad ds_L^2={1\over \pi\mu b^2}{dzd\bar{z}\over\left(1-|z|^{2}\right)^2}\propto ds_{\rm EAdS_2}^2\ .
\end{align}
Thus, in the $a\to 1$ limit, the space is Euclidean $AdS_2$.\footnote{In the strict $a\to 1$ limit, the normal derivative of the Liouville field at the boundary $\partial_r\phi\to\infty$, namely, the extreme opposite of the Neumann boundary condition. So the boundary condition essentially becomes Dirichlet. This suggests that the $E_L=0$ ground state corresponds to a ZZ-brane \cite{Zamolodchikov:2001ah}. Since $E_L\sim{\cal O}(b^3)\ll 1$, we may regard the Liouville dual of the JT gravity as excitations on ZZ-branes.} We will come back to this point later. 
The boundary length and area are given, respectively, by
\begin{align}\label{blength_area}
\ell =\int_{|z|=1} d\tau e^{b\phi}={1\over \kappa b^2}{2a\over 1-a^2}\ ,\qquad
A=2\pi\int_0^1rdre^{2b\phi}={1\over\mu b^2}{a^2\over 1-a^2}\ .
\end{align}
Note that $\ell\sim {\cal O}(1/b^3)$ that is consistent with Table \ref{table:correspondence}.
The saddle point value of the Liouville action can be calculated as
\begin{align}
S^{\rm saddle}_L={1\over b^2}\biggl[\underbrace{\frac{a^2}{1-a^2}}_{={\ell^2\over 4\pi A}-1}+\ln a
\underbrace{-{1\over 2}\ln(\pi\mu b^2)}_{{\rm counting}\,\,\chi({\cal M})}\biggr]+\mu A+\mu_B\ell\ .
\end{align}
Using $a\simeq 1-1/(\kappa\ell b^2)$, we can express the saddle point action in terms of $\kappa$, $\ell$ and $b$:
\begin{align}\label{saddleaction}
e^{-S^{\rm saddle}_L}=e^{{1\over 2b^2}\ln\mu-\kappa\ell +{1\over 2\kappa\ell b^4}+\cdots}\sim e^{S_0}K_{1/b^2}(\kappa\ell)
\end{align}
where the ellipses denote the terms that vanish in the $b\to 0$ limit and an unimportant constant. The classical action only reproduces the exponent of the modified Bessel function in an asymptotic expansion, but the prefactor can be reproduced by including the Gaussian fluctuations about the saddle point.
Observe that $1/(2\kappa\ell b^4)=2\pi^2\gamma/\beta$ according to Table \ref{table:correspondence} which agrees with the exponent of the JT disk partition function \eqref{JTdiskpartitionfunction}, whereas the factor $e^{-\kappa\ell}$ is precisely cancelled by the factor $e^{\kappa\ell}$ in \eqref{Laplace} which appears due to the shift of the Laplace conjugate variable of $\ell$ from $\mu_B$ to $-E_L=\mu_B+\kappa$.

Finally, we would like to add a side remark: If we consider the $\eta\ne 0$ case, one can easily find the relation
\be
\eta={1\over 2}\left(1-{\beta\over\pi\gamma}\sqrt{\gamma E\over 2}\right)
\ee
in the strict $b\to 0$ limit.
So $\eta=0$ implies that $E=2\pi^2\gamma/\beta^2$. This can be interpreted as the classical average energy:
\be
\langle E\rangle \equiv -\del_{\beta}\ln Z^{\rm disk}_{\rm JT}(\beta)\simeq 2\pi^2\gamma/\beta^2
\ee
in the classical regime $\gamma/\beta\gg 1$.

\subsection{The JT/Liouville geometric correspondence}
\label{Sec:LJTgeometry}

As promised, we now return to the geometry of the Liouville quantum gravity \eqref{LiouvilleCLgeometry}.
In the $a\to 1$ limit, the metric becomes
\begin{align}
ds_L^2={1\over 4\pi\mu b^2}{4dzd\bar{z}\over \left(1-|z|^2\right)^2}= {1\over 4\pi\mu b^2}ds_{\rm EAdS_2}^2\ ,
\end{align}
where the last EAdS$_2$ metric has the scalar curvature $R=-2$.
Thus if we wish to match the radii of the Louville and JT EAdS$_2$, we need to scale the Liouville gravity metric by
\be\label{rescaling}
ds_L^2\qquad\longrightarrow\qquad dS_L^2\equiv 4\pi\mu b^2ds_L^2=ds_{JT}^2
\ee
Note that this corresponds simply to a constant shift of the Liouville field $2b\phi\to 2b\phi_{JT}\equiv 2b\phi+\ln(4\pi\mu b^2)$.
Since $4\pi^2\ell\kappa b^4=\beta/\gamma$ according to Table \ref{table:correspondence}, 
if we rescale the Liouville gravity metric as in \eqref{rescaling}, the boundary length and its Laplace conjugate energy are rescaled to
\be\label{rescaledL}
\ell\,\,\longrightarrow\,\, L\equiv \sqrt{4\pi\mu}b\ell={\beta\over 2\pi\gamma}b^{-2}\ ,\qquad
E_L\,\,\longrightarrow\,\,{\cal E}_L={E_L\over \sqrt{4\pi\mu}b}=\left(2\pi\gamma E\right)b^2\ .
\ee
Since $a\simeq 1-1/(\kappa\ell b^2)= 1-2\pi/L<1$, the rescaled metric of the Liouville gravity is, more precisely, a cutoff EAdS$_2$:
\begin{align}\label{Liouvillespace}
dS_L^2={4a^2dzd\bar{z}\over \left(1-a^2|z|^2\right)^2}\ .
\end{align}
Transforming the coordinates by $z=e^{2\pi y\over\beta}$, the metric becomes a cutoff EAdS$_2$ black hole
\begin{align}
dS_L^2=\left({2\pi\over{\beta}}\right)^2{dyd\bar{y}\over \sinh^2\left({\pi(y+\bar{y})\over\beta}+\ln a\right)}
\simeq\left({2\pi\over{\beta}}\right)^2{dyd\bar{y}\over \sinh^2\left({\pi(y+\bar{y})\over\beta}-{4\pi^2\gamma b^2\over \beta}\right)}\ .
\end{align}
In the zero temperature limit $\beta\to\infty$ with $y=\sigma+i\tau$, the metric becomes the Poincar\'e EAdS$_2$ and we can identify the boundary cutoff 
with
\be
\sigma_{\rm cutoff}= (2\pi\gamma) b^2
\ee
which is proportional to the renormalized dilaton $\Phi_{\rm r}=2\pi\gamma$ as defined in \eqref{dilaton}. 
Alternatively, we can view $b^{-2}$ as the dilaton $\Phi$ at the cutoff surface since $b^{-2}=\Phi_{\rm r}/\sigma_{\rm cutoff}=\Phi(\sigma_{\rm cutoff})$.  
We add one more observation which provides a coherent picture: the rescaled boundary length $L$ in \eqref{rescaledL} is given, in terms of the boundary cutoff, by 
\be
L={\beta\over \sigma_{\rm cutoff}}\ .  
\ee
So $L$ is the proper boundary length at the cutoff surface $\sigma=\sigma_{\rm cutoff}$ in the AdS$_2$ space with the boundary circumference $\beta$ in agreement with the JT gravity. A summary of the correspondence for the rescaled parameters is listed in Table \ref{table:correspondence_rescaled}.
\begin{table}[h!]
\centering
\begin{tabular}{|c|c|c|}
\hline
  \multicolumn{3}{|c|}{JT gravity vs. Liouville CFT} \\
  \hline
&\quad JT gravity \quad&\quad Liouville CFT \quad\\\hline
genus counting & $S_0$ & $\ln\mu/(2b^2)$ \\\hline
dilaton & $2\pi\gamma$ & $b^{-2}=2\pi\gamma/\sigma_{\rm cutoff}$ \\\hline
boundary length & $\beta$ & $L=\beta/\sigma_{\rm cutoff}$ \\\hline
energy & $E$ & ${\cal E}_L=E\sigma_{\rm cutoff}$ \\\hline
\end{tabular}
\caption{The correspondence between the (rescaled) parameters of the two descriptions}\label{table:correspondence_rescaled}
\end{table}

\section{Replica Wormholes}
\label{Sec:RW}

The replica wormholes are the key to the existence of the islands that play a central role in a recent proposal for the resolution of the black hole information paradox \cite{Penington:2019kki, Almheiri:2019qdq}. They yield genuine gravitational contributions to the bulk entanglement entropy (EE) \cite{Lewkowycz:2013nqa} 
that goes beyond the Ryu-Takayanagi formula of minimal surface areas \cite{Ryu:2006bv, Ryu:2006ef}.
As the name suggests, they make an appearance in the replica trick for the computation of EE \cite{Calabrese:2004eu, Cardy:2007mb}.
In contrast to ordinary quantum field theories, in the presence of gravity, the replicated spacetimes can be connected by wormholes and, under certain circumstances, 
they give rise to dominant contributions to the entanglement entropy.

\subsection{A single cosmic brane}
\label{Sec:single}

\begin{figure}[h!]
\centering
\centering \includegraphics[height=2.3in]{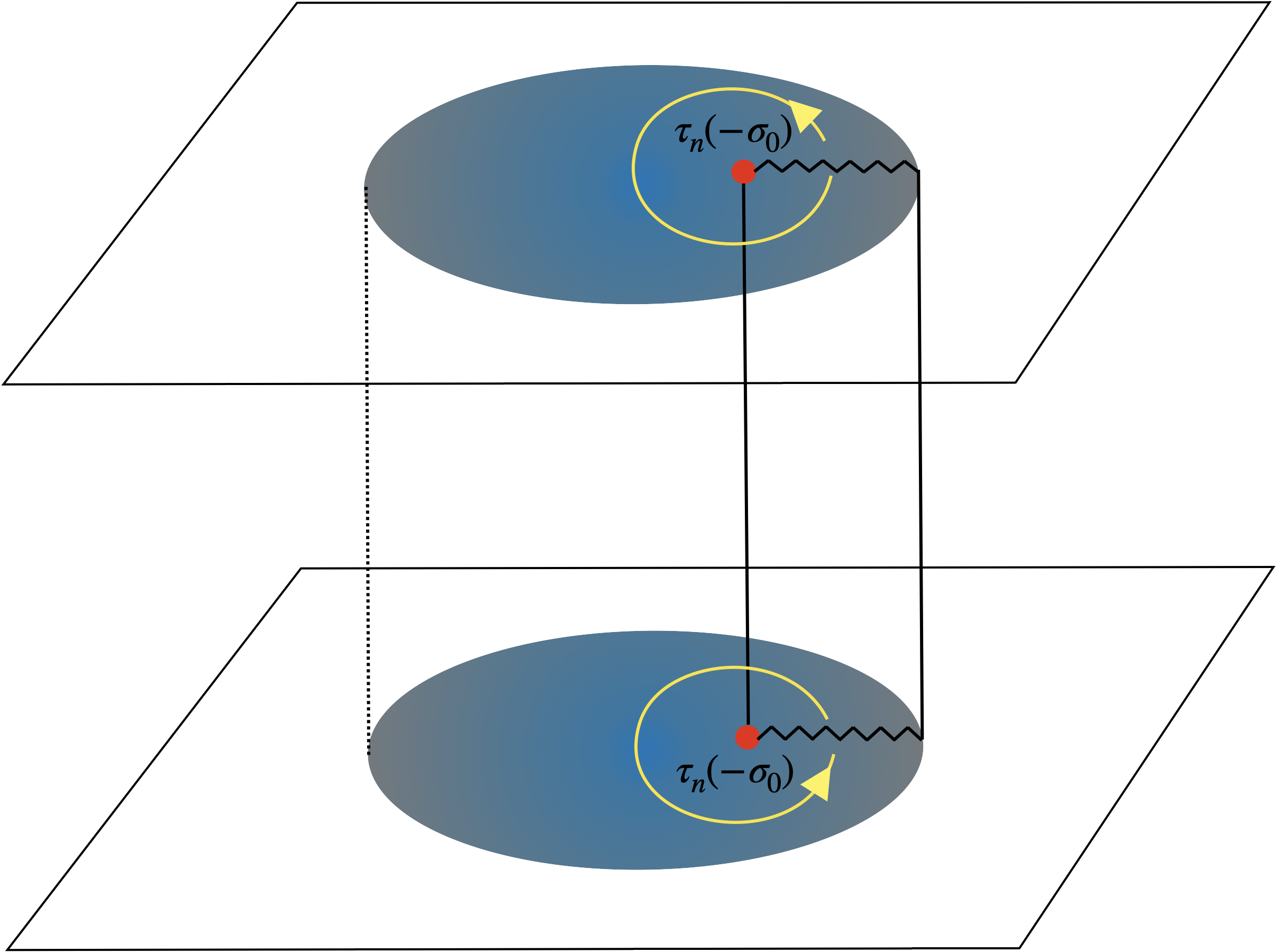}
\vspace{-0cm}
\caption{An illustration of replica wormholes: The dark blue disk is the 2$d$ Euclidean space $H_2=EAdS_2$. The space is $n$-tiply replicated and only two of them are shown.   
The red blob is the twist operator $\tau_n$ of the Liouville CFT that creates an $n$-th root branch cut (black wavy line) corresponding to a single ``cosmic brane'' that creates deficit angle $2\pi/n$ and will be located at the quantum extremal surface (QES), $y=\bar{y}=-\sigma_0$.}
\label{fig:single}
\end{figure}

The replica wormholes are created by co-dimension two cosmic branes which are the sources of conical singularities with deficit angle $2\pi/n$ \cite{Almheiri:2019qdq}.
In the JT gravity, the cosmic branes are point-like in two spacetime dimensions and so it is natural to identify them with the twist operators in the Liouville CFT description
as illustrated in Figure \ref{fig:single}. 

Our proposal is that the gravitational part of the bulk EE in the JT gravity can be described by the twist operators in the Liouville CFT:
\begin{align}\label{EEgravity}
S_{\rm G}\equiv S_0+S_{\rm dilaton}=-\lim_{n\to 1}{\partial\over\partial n}\Tr\rho_A^n
\quad\mbox{with}\quad
\Tr_{A}\rho^n=U(0)^{-1}\epsilon^{2\Delta_n}\langle\!\!\!\! \underbrace{\tau_n(z)}_{\rm cosmic\,\,brane}\!\!\!\!\rangle_{\rm FZZT}
\end{align}
where the twist operator $\tau_n(z)$ is composed of the Liouville, matter, and ghost sectors with a little ``twist''. As we argue in a moment, more explicitly, it takes the form
\begin{align}\label{taun}
\tau_n(z)=\left|d\zeta/dz\right|^{2\Delta_n}\!\!V_{\alpha_n}(\zeta)\,\tau^m_n(z)\tau^{bc}_n(z)\quad\mbox{with}\quad V_{\alpha_n}(\zeta)=e^{2\alpha_n\phi(\zeta)}\quad\mbox{and}\quad \zeta=az
\end{align}
where $\tau^m_n(z)$ and $\tau^{bc}_n(z)$ are the twist operators in the matter and $bc$ ghost sectors, respectively, and the conformal dimensions are
$\Delta_n=\bar{\Delta}_n= {c_L\over 24}\left(n-{1\over n}\right)$, $\Delta^m_n=\bar{\Delta}^m_n={c_m\over 24}\left(n-{1\over n}\right)$, and $\Delta^{bc}_n=\bar{\Delta}^{bc}_n={c_g\over 24}\left(n-{1\over n}\right)$. The central charges are $c_L=1+6Q^2$, $c_m=25-6Q^2$, and $c_g=-26$. 
A subtle but very important point is that the boundary for the Liouville twist operator $V_{\alpha_n}(\zeta)$ is at $|\zeta|=1$ as opposed to $|z|=1$. The reason is that as $n\to 1$ that is of our interest, from \eqref{ClLiouville}, the classical Liouville field becomes 
\begin{align}\label{Cltwistop}
e^{2\alpha_n\phi(z)}\propto\left({a^2\over (1-a^2z\zb)^2}\right)^{\alpha_n\over b}\qquad\mbox{where}\qquad
{\alpha_n\over b}\simeq \Delta_n\ .
\end{align}
This is essentially the bulk one-point function of the Liouville twist operator as $n\to 1$. So to be consistent with the semiclassical picture, the boundary for the Liouville twist operator looks as if it is at $|\zeta|=a|z|=1$ due to the nontrivial FZZT boundary condition. 
In the meantime, since the matter and ghosts are free fields, there is no such ``twist'' of the story. 
A few more remarks are in order: 
There are two possible values for $\alpha_n$. However, the appropriate choice turns out to be
\be
\alpha_n={Q\over 2}-\sqrt{{Q^2\over 4}-{1+6Q^2\over 24}\left(n-{1\over n}\right)}
\simeq {1\over 2b}\left(1-\sqrt{1-\left(n-{1\over n}\right)}\right)\ .
\ee
Note that $\lim_{n\to 1}\alpha_n=0$ and $\Tr\rho_A=1$ thanks to the factor $U(0)^{-1}$ since $U(0)\equiv \langle {\bf 1}\rangle_{\rm FZZT}$ as defined in \eqref{FZZT1pt}. The location $z$ will be at the quantum extremal surface (QES) upon the inclusion of the additional bulk matter sector.

As evidence for our proposal, we are going to show that our prescription reproduces the gravitational part of the bulk EE  \cite{Almheiri:2019qdq}:
\be\label{gravEE_AHMST}
S_G=S_0+{4\pi^2\gamma\over\beta\tanh{2\pi \sigma_0\over\beta}}\qquad\mbox{where}\qquad 2\pi\gamma=\Phi_{\rm r}\ .
\ee
First, it is easy to see how the $S_0$ piece can be reproduced. $\Tr\rho_A^n$ in \eqref{EEgravity} contains the factor
\begin{align}\label{mudep1pt}
U(0)^{-1}U(\alpha_n)\propto \mu^{-{\alpha_n\over b}}=\mu^{-{1\over 2b^2}\left(1-\sqrt{1-(n-1/n)}\right)}\ .
\end{align}
Upon differentiating it by $n$ and taking the limit $n\to 1$, this indeed yields ${1\over 2b^2}\ln\mu=S_0$ in the contribution to the bulk entanglement entropy.
So the remaining task is to demonstrate how the dilaton piece can be reproduced from \eqref{EEgravity}.
The basic idea is that it emerges from the position dependence of the bulk one-point function for the Liouville twist operator:
\be\label{1ptinz}
\left|{d\zeta\over dz}\right|^{2\Delta_n}\left\langle V_{\alpha_n}(\zeta)\right\rangle_{\rm FZZT}={a^{2\Delta_n}U(\alpha_n)\over (1-a^2z\zb)^{2\Delta_n}}\ .
\ee 
We further transform it to the EAdS$_2$ black hole coordinates by $z=e^{2\pi y/\beta}$:
\begin{align}\label{1ptiny}
 \left\langle V_{\alpha_n}(y)\right\rangle_{\rm FZZT}
 ={\left({\pi\over\beta}\right)^{2\Delta_n}U(\alpha_{n})\over\left(\sinh\left({\pi(y+\bar{y})\over\beta}+\ln a\right)\right)^{2\Delta_n}}\ ,
\end{align}
where we recall that $a\simeq 1-4\pi^2\gamma b^2/\beta$ and so $\ln a\simeq -4\pi^2\gamma b^2/\beta$. 
By setting $y=\bar{y}=-\sigma_0$ and including the matter and ghost sectors, this yields
\begin{align}\label{singleEE}
S_G=S_0+{4\pi^2\gamma\over\beta\tanh{2\pi\sigma_0\over\beta}}
+{c_L\over 6}\ln\left({\beta\over \pi\epsilon_L}\sinh{2\pi\sigma_0\over\beta}\right)
+{c_m+c_g\over 6}\ln\left({\beta\over \pi\epsilon}\sinh{2\pi\sigma_0\over\beta}\right)+C\ ,
\end{align}
where $C$ is some irrelevant constant that is a part of the factor $\lim_{n\to 1}U(0)^{-1}\partial_nU(\alpha_n)$ which does not depend on $S_0$, $\beta$, $\gamma$, or $\sigma_0$.
This can be absorbed into the UV regulator $\epsilon_L$ of the Liouville sector such that $e^{-6C/c_L}\epsilon_L=\epsilon$.
Since $c_{\rm tot}=c_L+c_m+c_g=0$, the last three terms cancel out, and as promised, this reproduces \eqref{gravEE_AHMST} and provides evidence for our proposal \eqref{EEgravity}.

\subsection{Two cosmic branes}
\label{Sec:double}

\begin{figure}[h!]
\centering
\centering \includegraphics[height=2.2in]{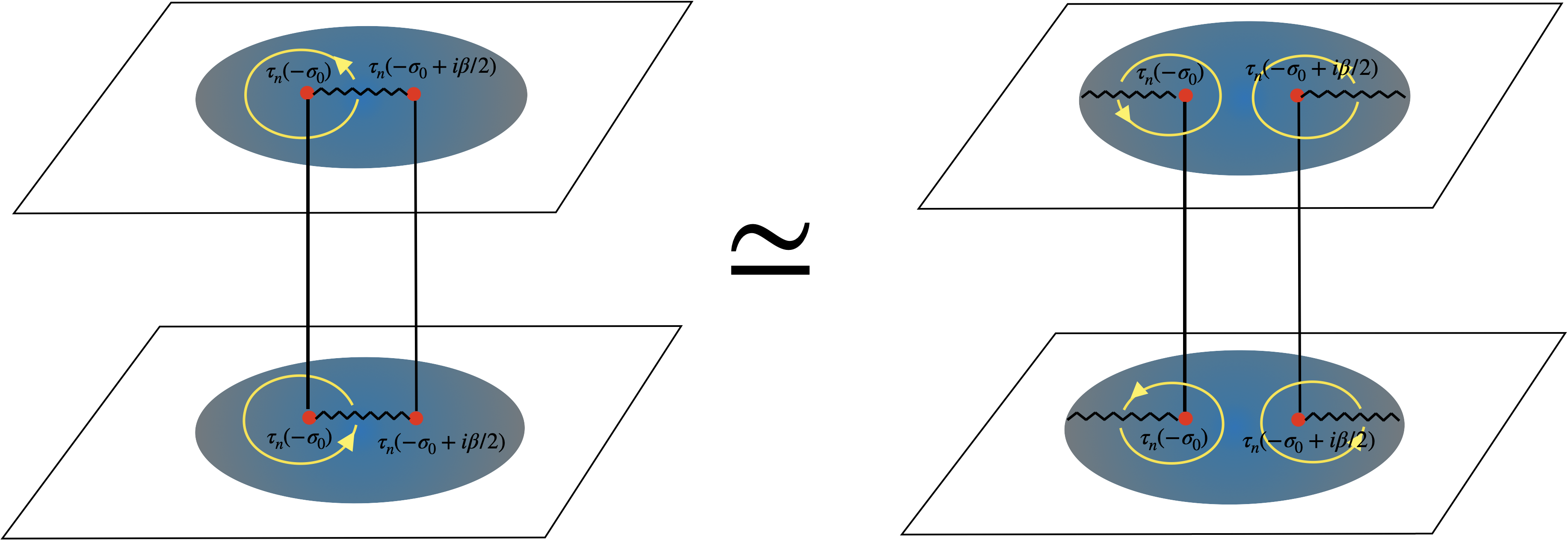}
\vspace{-0cm}
\caption{The replica wormholes in the eternal black hole: The dark blue disk is the 2$d$ Euclidean space $H_2=EAdS_2$. The space is $n$-tiply replicated and only two of them are shown. On each $H_2$, there is a pair of twist operators $\tau_{n}$ of the Liouville CFT (red blobs) corresponding to a pair of cosmic branes that create an $n$-th root branch cut (black wavy lines) between them (on the left), or alternatively, a pair of branch cuts emanating out to the boundary (on the right).}
\label{fig:double}
\end{figure}

In application to the information paradox in the eternal black hole, we need to consider the left-right symmetric two cosmic branes \cite{Almheiri:2019qdq}.
In the Liouville CFT description, this corresponds to the insertions of two twist operators as illustrated in Figure \ref{fig:double}.

It is then most natural to replace the bulk one-point function in  the proposal \eqref{EEgravity} with the bulk two-point function with the FZZT boundary condition:
\begin{align}\label{ReplicaWH2}
\Tr\rho_A^n=U(0)^{-1}\epsilon^{4\Delta_n}\langle \underbrace{\tau_n(z_1)\tau_n(z_2)}_{\rm two\,\, cosmic\,\,branes}\rangle_{\rm FZZT}\ ,
\end{align}
where $\tau_n(z)$ is defined in \eqref{taun} and the locations $z_1$ and $z_2$ of the cosmic branes are chosen in a left-right symmetric way and will be at the QES upon the inclusion of the additional bulk matter sector. 
Note that the RHS is properly normalized so that $\Tr\rho_A=1$, since the bulk two-point function becomes $U(0)=\langle{\bf 1}\rangle_{\rm FZZT}$ as $n\to 1$ despite the fact that we do not know the exact expression for the bulk two-point function.

As done in the case of a single cosmic brane, we are going to test our proposal against the result for the gravitational part of the bulk EE in the JT gravity  \cite{Almheiri:2019qdq}:
\be\label{gravEE_AHMST2}
S_G=2\left(S_0+{4\pi^2\gamma\over\beta\tanh{2\pi \sigma_0\over\beta}}\right)\qquad\mbox{where}\qquad 2\pi\gamma=\Phi_{\rm r}
\ee
which is simply twice the entropy for a single cosmic brane. At first, our proposal \eqref{ReplicaWH2} yields an extra contribution due to the interaction between the two cosmic branes. However, as we will show, the interaction becomes negligible at late times when the time evolution is taken into account, in agreement with \eqref{gravEE_AHMST2}. 

First, in a way similar to the case of a single cosmic brane, it is easy to see how the $2S_0$ piece can be reproduced from \eqref{ReplicaWH2}.
The $\mu$ dependence can be easily inferred from the Liouville action \eqref{LiouvilleAction}, or better \eqref{LiouvilleActionRescaled}, with operator insertions \cite{Zamolodchikov:1995aa}:
\begin{align}
U(0)^{-1}\langle V_{\alpha_n}(\zeta_1)V_{\alpha_n}(\zeta_2)\rangle_{\rm FZZT}\propto \mu^{-{2\alpha_n\over b}}\quad\mbox{with}\quad V_{\alpha_n}(\zeta)=e^{2\alpha_n\phi(\zeta)}
\end{align}
where $\zeta=az$.
Comparing it with \eqref{mudep1pt} for a single cosmic brane, this is simply twice and so it indeed yields $2S_0$ in the EE computed from \eqref{ReplicaWH2}.
The dilaton contribution once again comes from the position dependence of the bulk two-point function for the Liouville twist operators. For general bulk two-point functions, it is hard to find the exact dependence on the cross-ratio $\eta=|\zeta_1-\zeta_2|^2/|1-\zeta_1\bar{\zeta}_2|^2$. However, in the case of all twist operator correlators, the answer is known \cite{Calabrese:2004eu}:
\begin{align}
\langle V_{\alpha_n}(\zeta_1)V_{\alpha_n}(\zeta_2)\rangle_{\rm FZZT}
\propto\left({|\zeta_1-\zeta_2|^2\over |1-\zeta_1\bar{\zeta}_1||1-\zeta_2\bar{\zeta}_2||1-\zeta_1\bar{\zeta}_2|^2}\right)^{2\Delta_n}\ .
\end{align}
As alluded, we see that there are two types of contributions: (1) individual cosmic branes $|1-\zeta_i\bar{\zeta}_i|^{-2\Delta_n}$ $(i=1,2)$ and (2) the interaction of the two cosmic branes $\eta^{2\Delta_n}$.
The former precisely reproduces the dilaton contribution in \eqref{gravEE_AHMST2}, and the latter, at first, adds an extra contribution.
However, as mentioned above, we will demonstrate that the interaction becomes negligible at late times upon Lorentzian continuation.

To complete the computation, we perform the coordinate transformation $z\mapsto  y=\beta\ln z/(2\pi)$ to the EAdS$_2$ black hole as was done in \eqref{1ptiny} for a single cosmic brane. For the left-right symmetric cosmic brane configuration, setting $y_1=-\sigma_0$ and $y_2=-\sigma_0+i\beta/2$, we find that
\begin{equation}
\begin{aligned}
\left\langle V_{\alpha_n}(-\sigma_0)V_{\alpha_n}\left(-\sigma_0+{i\beta\over 2}\right)\right\rangle_{\rm FZZT}&\propto
\left({\left({2\pi\over\beta}\right)^2e^{-{4\pi \sigma_0/\beta}}a^2|2ae^{-{2\pi \sigma_0/\beta}}|^2\over |1-a^2e^{-{4\pi \sigma_0/\beta}}|^2
|1+a^2e^{-{4\pi \sigma/\beta}}|^2}\right)^{2\Delta_n}\\
&\simeq 
\frac{\left({\beta\over 2\pi}\sinh\left({4\pi \sigma_0\over\beta}\right)\right)^{-4\Delta_n}}
{\exp\biggl[4\Delta_n\biggl(\underbrace{\varepsilon\coth{2\pi \sigma_0\over\beta}}_{\rm individual}+\underbrace{\varepsilon\tanh{2\pi \sigma_0\over\beta}}_{\rm interaction}\biggr)\biggr]}
\end{aligned}
\end{equation}
where $\varepsilon=4\pi^2\gamma b^2/\beta$ and we have used $a\simeq 1-4\pi^2\gamma b^2/\beta$ for a small $b$. Including the matter and ghost sectors and calculating the entanglement entropy from \eqref{ReplicaWH2}, we obtain\footnote{As in the case of a single cosmic brane \eqref{singleEE}, there is an unimportant constant $C'$ coming from $U(0)^{-1}$ and the constant in the bulk two-point function. However, this can be absorbed into the UV regulator $\epsilon_L$ of the Liouville sector in a way similar to the case of a single cosmic brane.} 
\begin{equation}\label{twoCBEEE}
\begin{aligned}
S_G=\underbrace{2\left(S_0+{4\pi^2\gamma\over \beta\tanh{2\pi \sigma_0\over\beta}}\right)}_{\rm JT\,\,result}
+\underbrace{{8\pi^2\gamma\over \beta}\tanh{2\pi \sigma_0\over\beta}}_{\rm interaction}
+\underbrace{{c_{\rm tot}\over 3}}_{=0}\ln\left({\beta\over 2\pi\epsilon}\sinh{4\pi\sigma_0\over\beta}\right)
\end{aligned}
\end{equation}
As remarked above, at first, the interaction between the two cosmic branes appears to add an extra contribution to the JT gravity result \eqref{gravEE_AHMST2}. 
However, the comparison must be made at late times of the black hole evaporation process. In order to do so, we need to analytically continue our result to that in the Lorentzian black hole and take into account the time evolution of the entanglement entropy. Recall that the spaces of our interest are
\begin{align}
dS_L^2={4d\zeta d\bar{\zeta}\over (1-\zeta\bar{\zeta})^2}=\left({2\pi\over\beta}\right)^2{dyd\bar{y}\over\sinh^2\left({\pi(y+\bar{y})\over\beta}+\ln a\right)}\ ,
\end{align}
where $\zeta=az$ and $z=e^{2\pi y/\beta}$. The Lorentzian continuations are given by
\begin{align}
(z,  \bar{z})&\quad \xrightarrow{\rm Lorentzian}\quad (u, v)= (x+t, x-t)\ ,\\
(y, \bar{y})&\quad \xrightarrow{\rm Lorentzian}\quad (y^+, y^-)=(\sigma+\tau, \sigma-\tau)\ .
\end{align} 
The first metric is the AdS$_2$ black hole in the Kruskal coordinates and the second is the AdS$_2$ Schwarzschild black hole, and 
the two are mapped to each other by
\begin{align}
x+t = e^{{2\pi\over\beta}(\sigma +\tau)}\ ,\qquad x-t = e^{{2\pi\over\beta}(\sigma -\tau)}\ .
\end{align}
The cosmic brane 1 is located at $(x,t)=(x_0, t)$ and the cosmic brane 2 is at $(x, t)=(-x_0, t)$.
So upon Lorentzian continuation, we find 
\begin{align}
(\zeta_1, \bar{\zeta}_1)& \quad \xrightarrow{\rm Lorentzian}\quad (\zeta_1^+, \zeta_1^-)=\left(ae^{{2\pi\over\beta}(-\sigma_0 + \tau)}, ae^{{2\pi\over\beta}(-\sigma_0 - \tau)}\right)\ ,\\
(\zeta_2, \bar{\zeta}_2)& \quad \xrightarrow{\rm Lorentzian}\quad (\zeta_2^+, \zeta_2^-)=\left(-ae^{{2\pi\over\beta}(-\sigma_0 - \tau)}, -ae^{{2\pi\over\beta}(-\sigma_0 + \tau)}\right)\ ,
\end{align}
where $\sigma_0\ge 0$. Note that the overall minus sign in $\zeta_2^{\pm}$ is in agreement with the $\beta/2$ shift of the Euclidean time which maps the right black hole to the left one. A similar calculation to the Euclidean case yields
\begin{equation}\label{twoCBEE}
\begin{aligned}
S_G&=\underbrace{2\left(S_0+{4\pi^2\gamma\over \beta\tanh{2\pi \sigma_0\over\beta}}\right)}_{\rm JT\,\,result}
+\underbrace{{4\pi^2\gamma\over \beta}{\sinh\left({4\pi\over\beta}\sigma_0\right)\over \cosh\left({2\pi\over\beta}(\sigma_0+\tau)\right)\cosh\left({2\pi\over\beta}(\sigma_0-\tau)\right)}}_{\rm interaction}\\
&+\underbrace{{c_{\rm tot}\over 6}}_{=0}\ln\left(\left({\beta\over \pi\epsilon}\right)^2\sinh^2{2\pi\sigma_0\over\beta}{\cosh\left({2\pi\over\beta}(\sigma_0-\tau)\right)\cosh\left({2\pi\over\beta}(\sigma_0+\tau)\right)\over \cosh^2{{2\pi\over\beta}\tau}}\right)\ .
\end{aligned}
\end{equation}
Note that this reduces to the Euclidean result \eqref{twoCBEEE} at $\tau=0$. As promised, at late times $\tau/\beta\gg 1$, the interaction contribution becomes negligible 
and the entanglement entropy \eqref{twoCBEE} agrees with \eqref{gravEE_AHMST2}.
Physically, as time evolves, the two cosmic branes move away from each other to the future infinity of the left and right horizons.


\section{Marginal defect deformation}
\label{Sec:Mdd}

The Liouville CFT hosts a class of operators that are somewhat similar in their character to the twist operators discussed in the previous section. 
Except for the similarity in that both operators create deficit angles, this section is, by and large, independent from Section \ref{Sec:RW}.
These operators correspond to the conical defects in the JT gravity \cite{Mertens:2019tcm, Mertens:2020hbs, Turiaci:2020fjj, Okuyama:2021eju, Maxfield:2020ale}.
Instead of aiming for a comprehensive study of the conical defects, we focus on a very special defect, ``marginal defect'' as defined and elaborated below, which allows us to obtain the  exact result.

\begin{figure}[h!]
\centering
\centering \includegraphics[height=1.3in]{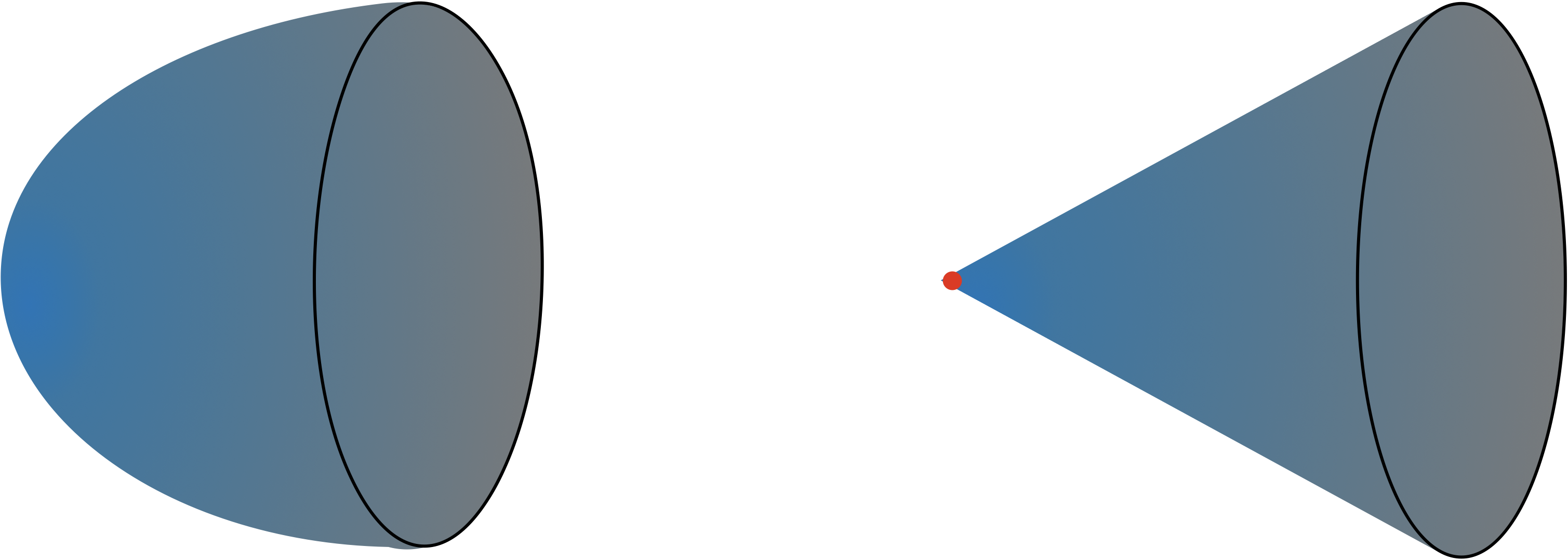}
\vspace{-0cm}
\caption{The conical defect: On the left is EAdS$_2$. An insertion of the defect operator ${\bf D}(z)$ at the origin $z=0$ creates a conical singularity and the resulting geometry is  EAdS$_2$ with a deficit angle.}
\label{fig:defect}
\end{figure}

To introduce and understand the operators of our interest, we first go back to the classical geometry discussed in Sections \ref{Sec:SCL} and \ref{Sec:LJTgeometry}.
From \eqref{ClLiouville} with the rescaling \eqref{rescaling}, the geometry created by an insertion of vertex operator $V_{\eta/b}(z)$ at the origin $z=0$ is given by
\begin{align}
dS_L^2={4a^2(1-2\eta)^2dzd\zb\over\left(|z|^{2\eta}-a^2|z|^{2(1-\eta)}\right)^2}
={4a^2dwd\bar{w}\over\left(1-a^2|w|^{2}\right)^2}\qquad\mbox{with}\qquad w=z^{1-2\eta}\ .
\end{align}
Going once around the origin $z\to e^{2\pi i}z$ yields $w\to e^{2\pi(1-2\eta)i}w$.  
One thus sees that the deficit angle $\delta=4\pi\eta$ is created around the operator $V_{\eta/b}(z)$.
In other words, the geometry is a (cutoff) EAdS$_2$ with a conical singularity as illustrated in Figure \ref{fig:defect}. 
Thus, in the range $0<\eta<1/2$, we can judiciously call these operators conical defects.

As discussed in \cite{Mertens:2020hbs}, the precise form of the CFT operator ${\bf D}(z)$ that describes a single defect in the JT gravity \cite{Mertens:2019tcm, Maxfield:2020ale}
is given by a marginal operator:\footnote{A convenient choice of the matter sector with $c_m=1-3(p-2)^2/p$ is a free field $X$ with the background charge $\alpha_0=(p-2)/\sqrt{2p}=\sqrt{Q^2-4}$ and the stress tensor $T=-{1\over 2}(\partial X)^2+i\alpha_0 \partial^2 X$. The vertex operator $V_X=e^{i\alpha X}$ has the conformal dimension $\Delta_m={1\over 2}\alpha(\alpha-2\alpha_0)$.}
\begin{align}\label{defectOP}
{\bf D}(z)=V_X\, e^{{2\eta\over b}\phi}\qquad\mbox{with}\qquad V_X\equiv e^{i\alpha X}
\end{align}
where $\alpha$ is determined by $\alpha(\alpha-2\alpha_0)/2+\eta(bQ-\eta)/b^2=1$ with $\alpha_0=\sqrt{Q^2-4}$ so that $V_XV_{\eta/b}$ is marginal.
Note, however, that in the partition function, the defect operators ${\bf D}(z)$ are inserted in an integrated form $\int d^2z{\bf D}(z)$ rather than as local operators.
As evidence for \eqref{defectOP}, it is easy to find that in the (semi)classical analysis for $\eta\ne 0$, the saddle point action \eqref{saddleaction} is generalized to \cite{Fateev:2000ik, Menotti:2006tc}
\begin{align}
e^{\kappa\ell}e^{-S^{\rm saddle}_L}=e^{{1-2\eta\over 2b^2}\ln\mu+{(1-2\eta)^2\over 2\kappa\ell b^4}+\cdots}=e^{(1-\delta/(2\pi))S_0+{2\pi^2\gamma\over \beta}(1-\delta/(2\pi))^2}
\end{align}
where we have used $\delta=4\pi\eta$. This indeed reproduces the classical part of the result found in \cite{Mertens:2019tcm, Stanford:2017thb}.

With this brief introduction to the Liouville CFT description of conical defects, in this paper, we would like to focus on a very special defect, namely, the $\eta=b^2$ case for which $\alpha=0$. This is nothing but the bulk cosmological constant operator $V_b=e^{2b\phi}$ of the Liouville theory. 
The deficit angle created by this operator is $\delta =4\pi b^2$. So in the JT limit $b\to 0$, naively, the deficit angle approaches to zero. For this reason, we call this operator a marginal defect. Nevertheless, as we will see, it yields a nontrivial and interesting effect on the disk partition function.

We consider a finite deformation to the Liouville and JT gravity actions  by the marginal defect:
\begin{align}\label{mdaction}
S_{\rm mdefect}=\lambda_L \int d^2ze^{2b\phi(z)}\ .
\end{align}
Using the results in Section \ref{Sec:DiskPT} and inverting the Laplace transform \eqref{Laplace}, the disk partition function with the marginal defect deformation reads
\begin{align}\label{defectdiskpartitionfunction}
Z^{\rm defect}_{\rm L}(\ell)&\simeq \sum_{k=0}^{\infty}{e^{S_0}\over k!}\int_0^{\infty}{dE_L\over 8\pi^4\kappa b^6}e^{-E_L\ell}
\biggl[{1\over 2\pi i}\int_{C-i\infty}^{C+i\infty}d\ell\, (-S_{\rm mdefect})^k\,e^{E_L\ell}\,{e^{\kappa\ell}\over\ell}K_{1/b^2}(\kappa\ell)\biggr]\ .
\end{align}
This can be calculated exactly either by a semiclassical method or a fully quantum mechanical calculation. Note that the semiclassical analysis suffices in the JT limit $b\to 0$.
We first use the semiclassical method which is more intuitive. Since the bulk cosmological constant term is nothing but the area of the Liouville geometry, from \eqref{blength_area}, one finds that
\begin{align}\label{mdactionvalue}
S_{\rm mdefect}=\lambda_L A\simeq \lambda_L{\kappa\ell\over 2\mu}\ .
\end{align}
Thus we have
\begin{equation}\label{mdefectPF}
\begin{aligned}
\rho^{\rm mdefect}_L(E_L)&\equiv {e^{S_0}\over 8\pi^4\kappa b^6}\sum_{k=0}^{\infty}{(-\lambda_L)^k\over k!}
{1\over 2\pi i}\int_{C-i\infty}^{C+i\infty}d\ell\, \left({\kappa\ell\over 2\mu}\right)^k\,e^{E_L\ell}\,{e^{\kappa\ell}\over\ell}K_{1/b^2}(\kappa\ell)\\
&= {e^{S_0}\over 8\pi^4\kappa b^4}\sum_{k=0}^{\infty}{1\over k!}\left(-{\lambda_L\kappa\over 2\mu}\right)^k
\del_{E_L}^k\rho_L(E_L)
= {e^{S_0}\over 8\pi^4\kappa b^4}\rho_L\left(E_L-\lambda_L\kappa/(2\mu)\right)\ .
\end{aligned}
\end{equation}
Adding a new entry to the dictionary by renormalizing the coupling 
\begin{align}
\lambda_L=8\pi^2 b^4\mu\lambda\ ,
\end{align}
the energy density in the JT gravity deformed by a marginal defect yields
\begin{align}\label{shiftedenergydensity}
\rho_{\rm JT}^{\rm mdefect}(E)dE=
\rho^{\rm mdefect}_L(E_L)dE_L=
e^{S_0}{\gamma\over 2\pi^2}\sinh\left(2\pi\sqrt{2\gamma (E-E_0)}\right)dE\ ,
\end{align}
where we have defined $E_0=\lambda/\gamma$. This coincides exactly with the energy density for generic defects at low temperature $\beta\gg 1$ found in \cite{Maxfield:2020ale}. Accordingly, the marginal defect action \eqref{mdactionvalue}, in terms of the JT gravity parameters, $S_{\rm mdefect}=\lambda\gamma/\beta$
is also the same as the one in \cite{Maxfield:2020ale}.\footnote{One of the main points in \cite{Maxfield:2020ale} is that this manifestly positive energy density \eqref{shiftedenergydensity} has bearing on the issue of negative energy density in the CFT dual of 3d pure gravity \cite{Maloney:2007ud, Keller:2014xba} pointed out by \cite{Benjamin:2019stq} and its resolution proposed in \cite{Benjamin:2019stq, Benjamin:2020mfz}. As explained in \cite{Maxfield:2020ale}, the first two terms of \eqref{shiftedenergydensity} in the small $\lambda$ expansion correspond to the contribution from the classical solutions and the fluctuations about them in a certain 2d/3d gravity and could become negative at low energies. However, the negativity of the energy density can be cured by summing over the contributions from off-shell conical defects, as the deformed energy density \eqref{shiftedenergydensity} indicates.} Even though this is a curious result, we have no clear understanding of if and why our result for the marginal defect, which corresponds to $\alpha_{MT}\to 1$ in \cite{Maxfield:2020ale}, should coincide exactly with their leading order result at low temperature for generic $\alpha_{MT}$.

As a double-check, we now provide an alternative calculation that is fully quantum mechanical. Since the marginal defect is the bulk cosmological constant, in the path integral, it can be expressed as $\mu$-derivatives:
\begin{align}
\partial_s\left\langle e^{-S_{\rm mdefect}}\right\rangle_{\rm FZZT}=\sum_{k=0}^{\infty}{\lambda_L^k\over k!}\partial_{\mu}^k\partial_sU(0)
=\sum_{k=0}^{\infty}{\lambda_L^k\over k!}
{\partial_{\mu}^k\over 2\pi i}\int_{C-i\infty}^{C+i\infty}d\ell\left(\mu^{1\over 2b^2}K_{1/b^2}(\kappa\ell)\right)F(\ell)
\end{align}
where function $F(\ell)$ collectively denotes the factors that are independent of $\mu$. By using the identity $\nu K_{\nu}(z)+zK'_{\nu}(z)=-zK_{\nu-1}(z)$ for the modified Bessel functions, it is straightforward to find that
\be
\del_{\mu}^k\left(\mu^{\nu/2}K_{\nu}(\kappa\ell)\right)=\left(-{\kappa\ell\over 2\mu}\right)^k\mu^{\nu/2}K_{\nu-k}(\kappa\ell)\ .
\ee
Since $\nu=1/b^2\to\infty$ in the JT limit, the net effect of an insertion of a single marginal defect is the factor $-\lambda_L\kappa\ell/(2\mu)$ in precise agreement with the classical area \eqref{mdactionvalue}. It thus reproduces the deformed energy density \eqref{shiftedenergydensity}.

In the broader context of the JT gravity/Liouville CFT correspondence, one may regard this discussion as an example of the exact deformation of the JT gravity that yields an interesting and physically sensible answer \eqref{shiftedenergydensity}.


\section{Discussions}
\label{Sec:Discussions}

The bulk CFT description may have technical advantages in analyzing the 2$d$ quantum gravity models based on the JT gravity and might potentially provide new insights into the black hole information paradox. For example, the Hilbert space structure of quantum gravity becomes clearer in this description since the Liouville CFT is an ordinary quantum field theory. In comparison to the path integral formulation of quantum gravity \cite{Colin-Ellerin:2020mva, Colin-Ellerin:2021jev}, it may be more straightforward in the Liouville CFT description to perform analytic continuation to Lorentzian spacetimes and it might help us understand better the real-time counterpart of replica wormholes.

In this paper, we have focused on the gravitational part of the bulk entanglement entropy in the JT gravity. However, this is only half the story and in order to study the black hole information paradox, we need to (1) add conformal matter radiation besides the $c_m=25-6Q^2$ matter that is a part of the gravity sector and (2) couple the Liouville CFT to a bath CFT on the flat space, via an interface, that represents the radiation in the asymptotically flat region of the spacetime \cite{Almheiri:2019psf, Almheiri:2019hni}.
One can hope that the standard CFT techniques might come in handy for the further expedition of the islands, beyond reproducing the Page curve, in order to uncover how black holes are processing the information in and out.
Even better, the Liouville CFT description, in the JT limit $b\to 0$, suggests that all can be done in semiclassical physics.

An important issue the JT gravity manifested itself is the factorization problem or an ensemble interpretation of AdS/CFT \cite{Saad:2019lba}.
This has bearing on the randomization of couplings via wormholes \cite{Coleman:1988tj, Klebanov:1988eh, Giddings:1988cx, Polchinski:1994zs} 
even though the connection may not be direct. A useful tool to approach the factorization problem can be provided by the Liouville CFT description 
of multi-boundary partition functions \cite{Mertens:2020hbs}. (See also a work \cite{Betzios:2020nry} which addressed the factorization problem, in particular, in the $c=1$ string theory dual to a unitary matrix quantum mechanics.) The Liouville CFT operators in the principal series $\eta=1/2+iP$ $(P\in\mathbb{R})$, 
the FZZT and ZZ-brane boundary states, that we have not explored in this paper, are all essential ingredients in this correspondence. 
We hope to address some of these issues in the near future.


\section*{Acknowledgments}

We are very grateful to the Nagoya University String Theory Group for giving us the opportunity
for regular meetings that motivated us to initiate this project. 
We would, especially, like to thank Yuki Miyashita, Tadakatsu Sakai, Yuki Sato, and Masaki Shigemori for discussions.
The work of SH was supported in part by the National Research Foundation of South Africa
 and DST-NRF Centre of Excellence in Mathematical and Statistical
 Sciences (CoE-MaSS).  Opinions expressed and conclusions arrived at are
 those of the authors and are not necessarily to be attributed to the NRF
 or the CoE-MaSS. The work of T. K. was supported by 
JSPS KAKENHI Grant Number JP19K03834.

\appendix


\end{document}